# Modulated Martensite: Why it forms and why it deforms easily

S. Kaufmann[1,2], R. Niemann[1,2], T. Thersleff[1], U. K. Rößler[1], O. Heczko[3,1], J. Buschbeck[1], B. Holzapfel[1,2], L. Schultz[1,2] and S. Fähler [1,2*]

[1] *IFW Dresden, PO Box 270116, 01171 Dresden, Germany*

[2] *Institute for Solid State Physics, Department of Physics, Dresden University of Technology, 01062 Dresden, Germany*

[3] *Institute of Physics, Academy of Science, Na Slovance 2, CZ-182 21 Praha 8, Czech Republic*

*corresponding authors email: s.faehler@ifw-dresden.de

**Abstract**

Diffusionless phase transitions are at the core of the multifunctionality of (magnetic) shape memory alloys, ferroelectrics and multiferroics. Giant strain effects under external fields are obtained in low symmetric modulated martensitic phases. We outline the origin of modulated phases, their connection with tetragonal martensite and consequences for their functional properties by analysing the martensitic microstructure of epitaxial Ni-Mn-Ga films from the atomic to macroscale.

Geometrical constraints at an austenite-martensite phase boundary act down to the atomic scale. Hence a martensitic microstructure of nanotwinned tetragonal martensite can form. Coarsening of twin variants can reduce twin boundary energy, a process we could observe from the atomic to the millimetre scale. Coarsening is a fractal process, proceeding in discrete steps by doubling twin periodicity. The collective defect energy results in a substantial hysteresis, which allows retaining modulated martensite as a metastable phase at room temperature.

In this metastable state elastic energy is released by the formation of a 'twins within twins' microstructure which can be observed from the nanometre to millimetre scale. This hierarchical twinning results in mesoscopic twin boundaries. Our analysis indicates that mesoscopic boundaries are broad and diffuse, in contrast to the common atomically sharp twin boundaries of tetragonal martensite. We suggest that observed extraordinarily high mobility of such mesoscopic twin boundaries originates from their diffuse nature which renders pinning by atomistic point defects ineffective.

**Contents:**

# 1. Introduction

Often martensitic microstructures appear more like modern art than physics. While the rigorous mathematical description of these complex microstructures is art on its own, there are only few cases where a nontrivial martensitic microstructure can be illustrated in an intuitive way. This hampers the understanding of martensitic materials like magnetic shape memory alloys [1], ferroelectrics [2] and other multiferroics [3], since their multiscale microstructure is crucial for their functional properties.

To understand different types of twinned martensitic microstructures we use epitaxial films made from Ni-Mn-Ga magnetic shape memory alloy as a model system. This archetypical ferromagnetic Heusler alloy undergoes a martensitic phase transition which produces a microstructure with twin variants of different crystallographic orientation. The orientation is switchable by magnetic field or stress [1]. Therefore, Ni-Mn-Ga can be classified as a ferromagnetic-ferroelastic, i.e. multiferroic, material. An appropriate twinned microstructure is essential to obtain a magnetic-field-induced ferroelastic reorientation process. In particular, an exceptionally high twin boundary mobility is a crucial prerequisite for magnetic shape memory alloys to reach the outstanding high strains close to 10% in moderate magnetic fields [4].

A martensitic transition affects the material from the atomic to the macroscale. We will show that geometrical rules derived from matching lattices connect all these length scales. This allows addressing of some fundamental questions about the relation between martensite microstructure and extraordinary functional properties in the modulated phases identified as adaptive martensites:

1) How do continuum concepts of martensite change at length scales where only a discrete number of unit cells are involved?
2) What decides between a martensitic microstructure consisting of a hierarchy of 'twins within twins' or branching of twin variants?
3) What consequences have these microstructural effects for phase stability and hysteresis?
4) What is the connection between the lattice at the atomic scale and the macroscopically observable shape changes mediated by microstructural process of easily moved twin boundaries?

We will sketch answers to these questions, but also point to further experiments required to confirm the suggested concepts. Therefore we will apply continuum descriptions of martensite, in particular the concept of adaptive martensite [5], branching of twin boundaries [6], hierarchy of twins [7] and fractal martensite [8]. In order to obtain a quantitative, but still descriptive understanding, first the geometrical constraints during a martensitic transition are reviewed. The experiments presented have been performed on epitaxial films. We are well aware that films differ from bulk in some aspects. In a separate chapter (7) we analyse these differences and show why films are beneficial for the present analysis. Finally the similarities to other modulated phases (in particular in ferroelectrics) are discussed to illustrate the universality of our conclusions.

## 2. Geometry of a martensitic transition

As a starting point for the present experiments it is sufficient to consider diffusionless transformations from a cubic austenite single crystal to a tetragonal martensite (assigned as NM). As sketched in Fig. 1 (a), there are three equivalent ways to deform the cubic unit cell to a tetragonal one. This results in three possible alignments of the tetragonal $c_{NM}$-axis with respect to the original austenite cube axes. Without loss of generality we consider the $c_{NM}$-axis to be longer than the austenite lattice parameter $a_A$. As the volume during the martensitic transformation remains almost constant, both $a_{NM}$-axes are shorter than $a_A$. In the absence of external fields or loads, none of the three possible orientations of the tetragonal martensite unit cell should be favoured. A uniform distribution is realized by martensite variants with different orientation where one variant consists of neighbouring unit cells with identical orientation.

The transformation from cubic to tetragonal lattice structure takes place by a ferroelastic first-order phase transition. Hence, during the transformation a phase boundary between the two lattices must appear. This interface between parent austenite and martensite phase is called habit plane. At one side of the habit plane the macroscopic extension is fixed by the austenite lattice constant $a_A$. This constrains the formation of martensite variants on the other side of the habit plane. Since the overall number of unit cells remains constant during a diffusionless transformation, the formation of only one martensitic variant with a lattice parameter different from the austenite would require huge elastic energy. This energy can be efficiently reduced if the martensitic variants are arranged with long $c_{NM}$-axis and short $a_{NM}$-axis alternating along the habit plane. Differently aligned martensitic variants are connected by twin boundaries, since these highly symmetric boundaries exhibit lower interface energy compared to other possible interfaces like ordinary grain boundaries.

The first mathematical description of this concept was given by Wechsler, Liebermann and Read [9] and independently by Bowles and McKenzie [10]. Although this phenomenological theory of martensite uses the crystallographic lattice constants of both phases to predict the orientation of the habit plane (see [11] for a modern mathematical description), it is ultimately a continuum model. Despite giving the fraction of the different martensitic variant widths, it cannot predict their absolute widths. Khachaturyan et al. [5] considered the case when the elastic energy due to the lattice misfit at the austenite-martensite interface is high compared to the twin boundary energy. In this case the overall energy can be minimized by decreasing the variant size down to only few atomic layers resulting in a regular twinning pattern on the nanometre scale. This (micro-) structure is described as adaptive martensite phase.

The adaptive phase forms in such a way, that the habit plane separating austenite and adaptive martensite is an exact interface. From the mathematic point of view this means that one eigenvalue of the austenite-martensite strain transformation matrix is exactly equal to one, Ref. [11], chap.7.1. Hence one of the (pseudo-orthorhombic) martensitic lattice constants is equal to the austenite lattice constant $a_A$. Volume conservation then requires that the other two eigenvalues are smaller and bigger than one, respectively [11]. In this special case the interface between austenite and martensite can be formed with one single variant of the martensite and no twinning is required. Though an exact habit can still be highly indexed, it represents a condition favourable for low hysteresis [30,31].

In order to illustrate the martensitic phase formation at the habit plane, the geometry for the model system Ni-Mn-Ga is exemplarily sketched in Fig. 1 (b). Here, we focus on the 14M modulated phase (also called 7M). Recently we could show that the 14M martensite is a nanotwinned adaptive phase, which consists of tetragonal building blocks of the non-modulated (NM) martensite [12], leading to an exact habit plane between 14M and austenite. One of the 24 possible habit plane orientations is sketched in Fig. 1 (b). Blue colours are used to illustrate two orientations of the tetragonal martensite cell ($c_{NM}/a_{NM}$ =1.23), the parent austenite is marked in red. The edge length of the building blocks in Fig. 1 is half the lattice parameter of the Heusler unit cell. Required by the diffusionless transformation, the number of austenite and martensite unit cells on each side of the habit plane is identical along the vertical direction. For the lattice constants of the Ni-Mn-Ga alloy this condition is approximately fulfilled if the width of variants having their $a_{NM}$-axis in vertical direction ($a_{NM}$-variants) is five building blocks and the one of $c_{NM}$-variants is two blocks. Then the apparent lattice parameter in vertical direction (assigned as $b_{14M}$) of this periodic arrangement of simple tetragonal unit cells has almost the same length as the identical number of austenitic unit cells. This nanotwinned lattice represents the most finely twinned periodic microstructure exhibiting an exact habit plane. Due to the discrete size of the involved building blocks, a further reduction is not possible. In structural analysis, this nanotwinned adaptive lattice seems to have a (pseudo-) orthorhombic symmetry, which is a lower symmetry compared to its tetragonal martensitic building blocks. A closer look at the structure reveals that the symmetry is further lowered by a small monoclinic distortion.

As described by Khachaturyan et al. [5] elementary geometry gives the twinning periodicity $d_1/d_2 = (a_{NM}-a_A)/(a_A-c_{NM})$. For the present sample, the measured lattice constants yield $d_1/d_2 = 0.417$ [12], which is quite close to the ideal value for a $(5\bar{2})_2$ modulation of $2/5 = 0.4$. This nanotwinned lattice is commonly described as a 7M modulated martensitic phase [16].

Due to the integer number of unit cells involved, the $d_1/d_2$ ratio for an ideal 7M modulation differs from the one expected from the lattice constants. Hence a low number of stacking faults needs to be inserted within the lattice to adapt the modulated lattice to the parent austenite. This agrees with crystallographic studies using diffraction data, describing this structure as incommensurately modulated phase [13].

Considering the symmetry of the ordered $L2_1$ Heusler structure of Ni-Mn-Ga, 14 building blocks are needed in order to start and finish with the same kind of atom. Hence the crystallographic correct description is 14M. Accordingly, the volume of the building blocks used in Fig. 1 is just one eighths of a Heusler unit cell. Compared to a unit cell the building block concept is more suitable to describe some features (in particular coarsening). Though in the following lattice constants of course refer to unit cells, it is more intuitive to think of building blocks.

However, the use of 5 and 2 building block thick variants is only one possible solution since, from the macroscopic point of view, all variant lamellae being multiples of this smallest nanotwinning would also provide an invariant plane strain. Indeed, X-ray diffraction (XRD) data of the sample examined here revealed that, in addition to the adaptive nanotwinned 14M martensite, there are also macroscopic variants of the tetragonal NM martensite. These macroscopic tetragonal variants have the same spatial orientation as the ones being only two or five unit cells thick [12]. The coexistence of both phases can be understood when considering annihilation of twin boundaries as transition mechanism between 14M and macroscopic NM. Such mechanism is described by Kohn and Müller [6] as a branching of twin boundaries when approaching an invariant interface (habit plane) (Fig. 2). This reduces the elastic energy originating from the difference of lattice constants on cost of increasing twin boundary energy. We reformulate this approach from a different point of view. Starting from the habit plane, where the twin variants are only few atomic layers thick, twin boundaries can annihilate with increasing distance to the habit plane in order to reduce twin boundary energy. We will refer to this process, which is sketched in Fig. 1 (b to d), as coarsening of twin variants, also suggested in [5].

The habit plane between austenite and 14M is close to a $(101)_A$ plane [12]. As its orientation is determined by macroscopic constraints, it usually deviates from a low indexed plane. This results in a significant disturbance at the atomic scale, in contrast to a twin boundary, which is atomically sharp. In Fig. 1 (b) these lattice defects are symbolized by the diffuse grey region, which can expand over several atomic planes.

For the present microstructural analysis, we use the same epitaxial film, for which we could prove the adaptive nature of the 14M martensite by integral XRD methods [12]. To illustrate the film architecture, Fig. 1 can be also regarded as a sketch of the film cross section. The substrate would be aligned vertically left of Fig. 1 (b) and connected to the austenite. The residual austenite at the substrate interface cannot transform, since the rigid substrate hinders any length changes. This frozen phase transition is the crucial precondition for the following analysis.

## 3. Experimental

Epitaxial Ni-Mn-Ga films were deposited on MgO(001) using DC-magnetron sputtering [14] at a deposition temperature of 250°C (sputter power = 100 W, base pressure = $10^{-9}$ mbar, working pressure = 8 x $10^{-4}$ mbar). The composition of the film was $Ni_{54.8}Mn_{22.0}Ga_{23.2}$ as determined by electron dispersive X-ray (EDX) spectroscopy using a stoichiometric $Ni_2MnGa$ standard (error of less than 0.5 at.%). The structure was studied by X-ray diffraction. At room temperature the film consists of residual austenite with $a_A$ = 0.578 nm, non-modulated martensite NM with $a_{NM}$ = 0.542 nm, $c_{NM}$ = 0.665 nm and modulated 14M martensite with $a_{14M}$ = 0.618 nm, $b_{14M}$ = 0.578 nm and $c_{14M}$ = 0.562 nm. All lattice parameters are described with respect to the ordered $L2_1$ Heusler unit cell. Detailed structural analyses of this film [12] and similar films with only 14M modulated structure [15] have been published elsewhere.

A cross section of the film was prepared by focused ion beam cutting (FIB Crossbeam 1540 XB, Zeiss) and revealed a film thickness of about 420 nm. The film surface was examined by Atomic Force Microscopy (AFM) in tapping mode (Digital Instruments Dimension 3100) and high resolution Scanning Electron Microscopy (SEM) using a backscattered electron detector (LEO 1530 Gemini).

## 4. Microstructure of 14M and NM martensite

The structural analysis of the epitaxial Ni-Mn-Ga film revealed the coexistence of austenite, tetragonal martensite (NM) and 14M modulated martensite [12]. Here, the microstructure is investigated at different length scales by SEM including FIB-cuts for cross section analysis and AFM in order to assign the different features in microstructure to the phases and their structure. These measurements benefit from the epitaxial relationship between film and substrate. The austenitic Ni-Mn-Ga unit cell is rotated by 45° with respect to the MgO<100> substrate edges. In top view pictures (Fig. 3 and Fig. 4), the substrate edges are parallel to the picture frame.

When examining the film surface on a large scale of several tens of micrometers, SEM micrographs (Fig. 3 (a) and (b)) reveal two different patterns in the twinned microstructure with characteristic features. The first pattern shows periodic wavy features which are rotated by about 45° with respect to the substrate edges (marked by the green line in Fig. 3 (b)). The second type of patterns shows straight and more regular features perpendicular or parallel to the substrate edges (red line in Fig. 3 (b)). These twin boundaries are not curved and their periodicities vary from a few tens of nanometres up to several micrometres. In the following, we will identify the first pattern as twinned regions consisting of 14M modulated martensite (green) while the second patterns will be identified as twinned non-modulated (NM) martensite (red).

While at the film surface only the traces of the twin boundaries can be observed, the combination with a FIB cut of the sample allows to determine the spatial orientation of the twinning planes (bottom of Fig. 3 (c)). Twin boundaries in bulk Ni-Mn-Ga are {110} planes [16] and indeed both twin boundary orientations can be identified with differently aligned {110} planes. For the present thin film, however, not all six possible orientations occur equally for both structures. As sketched in Fig. 3 (d) 14M twin boundaries are inclined by 45° towards the substrate plane, whereas NM twin boundaries are perpendicular to the substrate. This shows that the substrate constraint selects a certain twinning microstructure as it breaks the symmetry between the different primary twinning modes and habit planes that are equivalent in a bulk single crystal. However, the microstructure of the epitaxial film still reflects the fourfold symmetry induced by the (001) oriented cubic substrate.

In order to quantitatively assign the features on the film surface to the respective martensite phases, we start from the smallest feature size of the first pattern, as observed in the area around the green line in Fig. 3 (b). On average, these wavy features run along lines under 45° with respect to the substrate edges and exhibit a periodicity of about 80 nm. The meandering of these patterns indicates a high stress state since bending of a twin boundary increases its length and, thus, its energy. In one of these regions the surface topography is examined by AFM (Fig. 4 (a)). The image reveals a periodic triangular height profile with a characteristic topography angle of $\alpha = 5.5°$ (green line in Fig. 4 (b)). Since the angle between two martensitic variants connected by a {101} twin boundary is determined by the martensitic lattice constants by $c/a = \tan(45°-\alpha/2)$, it can be used for phase identification [17]. When considering all possible combinations of martensitic lattice constants in the sample measured by XRD, the $c/a$ of 0.909 obtained from surface topography can only be caused by a $c_{14M}$-$a_{14M}$ twin boundary ($c_{14M}/a_{14M} = 0.91$). This allows identifying this pattern with twin boundaries between (pseudo-)orthorhombic variants which share their $b_{14M}$-axis directions. Furthermore, these features (orientation, meander like, topography angle, spacings) are identical to the ones observed in a film of identical thickness but different composition, which only consists of 14M martensite according to XRD [17]. Hence we can identify the phase in this finely twinned area as 14M.

The advantage of this indirect method of local phase identification is that one also obtains the orientation of the 14M unit cell with respect to the twin boundaries. As lines on the film surface are traces of $c$-$a$-twin boundaries of the 14M martensite, $c_{14M}$ and $a_{14M}$ are alternately directed perpendicular to the film surface (or parallel to the green line in Fig. 4 (a), respectively). Accordingly the third orthorhombic axis, $b_{14M}$, must always be aligned in the film plane and parallel to the twin boundary (or perpendicular to the green line in Fig. 4 (a)). We can exclude a significant amount of $a$-$b$ and $b$-$c$ twin boundaries of the 14M martensite, as no twinning angle corresponding to such ratios of lattice parameters is observed at the film surface.

Patterns exhibiting lines parallel to the substrate edges (Fig. 3 (a) and (b)) have not been observed in the pure 14M films [17] but only in films revealing NM martensite by XRD [18]. This suggests that these features originate from macroscopic NM twin variants. To illustrate the twin boundary orientation, one may use the sketch in Fig. 1 (c), considering the paper plane as the film plane and the epitaxial relationship, which results in a rotation by 45° around the substrate normal. Thus, the orientation of twin boundaries corresponds to a NM microstructure, where $a_{NM}$ and $c_{NM}$ alternate from variant to variant in the film plane (the second $a_{NM}$ is always pointing out-of-plane). Since these straight lines are traces of twin boundaries penetrating the film surface, $a_{NM}$ - $c_{NM}$ twin boundaries are aligned perpendicular to the substrate (red plane in Fig. 3 (d)). Hence only two of the six possible {110} twinning planes are apparently realized in this phase. The origin of the symmetry breaking effect is discussed in chapter 7.

## 5. Coarsening of martensitic variants

Up to this point we have treated 14M and NM as different martensitic phases. The concept of adaptive martensite, however, implies that both, 14M and NM, have the same crystallographic structure. The only difference between both is the density of twin boundaries [12]. Following this concept, it is more appropriate to speak about 14M and NM having a different scale of microstructure instead of being different phases.

Following the concept of Kohn and Müller [6], branching of twin boundaries should occur when approaching the habit plane (Fig. 2.). This reduces the elastic energy originating from the difference of lattice parameters on cost of increased twin boundary energy. At each distance to the habit plane a regular twinning period forms within planes being parallel to the habit plane. Within each plane the ratio of both variant lengths has to remain constant as this minimizes the elastic energy. To achieve this microstructure, some bending of the twin boundaries is unavoidable when approaching the habit plane (Fig. 2).

As sketched in Fig. 1 (b) and (d), we suggest that coarsening of variants connects the 14M and the NM microstructure. A first direct evidence for coarsening of martensitic variants arises from the micrograph of the film cross section (bottom of Fig. 3 (c)). When starting at the surface in a region identified as the NM martensite (red lines) one observes that the frequency in contrast doubles when approaching the interface to the rigid substrate surface. The expected further doublings when approaching the substrate are below the resolution of the SEM.

A quantitative analysis of the coarsening mechanism is possible by investigating the NM areas on the film surface. We analyzed variant fractions and periodicity of differently aligned NM variants using the SEM micrographs (see supplementary material online for more details). From the measured widths $d_1$ and $d_2$ of $a_{NM}$ and $c_{NM}$ variants, we calculated their ratio $d_1/d_2$ and wavelength $\Lambda = d_1 + d_2$, as summarized in Fig. 5. Additionally the values for a $(5\bar{2})_2$-stacked nanotwinned NM martensite (= 7M) as determined by XRD are shown. The ratios of the twin widths in all analyzed regions are always close to the value of $d_1/d_2 = 0.417$ expected from the NM lattice constants. These data show that the macroscopic length conservation determines the twin variant width ratio although the periodicity itself changes over several orders of magnitude. Obviously, the required compatibility of the martensite with the remaining austenite acts as macroscopic constraint that completely determines the characteristic lengths ratios in the twinned NM microstructure from the atomic up to micrometer scale.

The absolute size of the NM twin variants differs significantly in different regions in the sample. This indicates that various stages of the coarsening process are present within the sample. It seems that kinetic barriers allow for a quite wide spectrum of different metastable microstructures, i.e. the coarsening can stop due to some kinetic reason in different incompletely coarsened stages.

As the martensitic transition proceeds from austenite over 7M to NM martensite we suggest the following scenario for the coarsening process. According to the concept of adaptive martensite we expect 7M in the vicinity to the habit plane, since elastic energy is minimized by the smallest twin variant size consisting of five and two building blocks, respectively. An increased variant size, however, would allow reducing the number of twin boundaries, which decreases twin boundary energy. For this coarsening process there are two constraints on the variant distribution in planes parallel to the habit plane. First, the length should remain constant, which minimizes the elastic energy. Second, the number of building blocks should remain constant, which reflects the fact, that the coarsening process is diffusionless. As illustrated in Fig. 1, both conditions are fulfilled, when the period is doubled (from (b) to (c)). In the following thought experiments it is illustrated why this is the simplest way to reduce energy. Reorienting just one unit cell from $a_{NM}$ to $c_{NM}$ (or vice versa) varies length (by the difference of both lattice constants) and is hence increasing elastic energy. Shrinking e. g. one *c*-variant by one building block and increasing another *c*-variant one accordingly, keeps the length constant. However, this process would leave the number of twin boundaries constant, and thus there is no driving force available for this process. Also a simple parallel translation of two neighbouring twin boundaries does not reduce twin boundary energy. Moreover, defects in the crystal structure associated with this process are expected to hinder this movement. Doubling of period is the simplest process to reduce twin boundary energy in a system with a discrete number of building blocks. More complex processes (e. g. annihilation of every second pair of twin boundaries) are possible, but unlikely since they break symmetry. In contrast to a statistical process coarsening is hence expected to be a coupled, collective mechanism. This scenario suggests that each periodic pattern observable at the film surface can be assigned to an integer generation $n$ of the coarsening.

The present experiments allow probing this concept experimentally by analysing the twinning wavelength $\Lambda = d_1 + d_2$. Starting point as the $0^{th}$ generation is 7M with a period length $\Lambda_0$ (see fig. 1 b). Elementary geometry and XRD data give $\Lambda_0 = a_{Nm} * c_{NM} / \sqrt{a_{NM}^2 + c_{NM}^2} * 7/2 = 1.471\,\text{nm}$ [12]. For all following generations $\Lambda$ doubles, thus period lengths should follow $\Lambda_n = 2^n * \Lambda_0$. This prediction can be tested by plotting $\Lambda$ values obtained for all regions available to analysis against integers $n$ that are assigned as coarsening generation of the different twin microstructures (Fig. 5). Since all data points are close to intersections between the $\Lambda_n$-line and integer values of $n$, we believe that the concept provides a valid quantitative description of the coarsening process. The combination of microscopic observations with structural analysis enables to track the mechanism of coarsening over three orders of magnitude. As it starts from the nanotwinned 7M, the lattice periodicity (modulation) fixes an absolute minimum length scale for the microstructure. Then, the coarsening proceeds by doubling variant width up to the micrometer regime. Thus, the coarsening also determines the widths of the macroscopic NM twin variants. This is in contrast to the common continuum approach of branching, where no minimum length scale exist. For continuum theory the length scale of twinning is determined by a macroscopic energy balance between interface energies and elastic stresses. Continuum theory would allow any rational multiple of wavelength to maintain an invariant length. However, this is not possible if a discrete number of tetragonal unit cells as elementary building blocks is involved. In that case variants can grow, i.e. coarsen, by doubling their widths. Hence coarsening of the adaptive martensite is discretized by the finite size of the atomic building blocks, an aspect that is commonly not considered during branching.

## 6.     Coarsening, branching and fractals

The present experiments seamlessly connect coarsening of a discrete number of building blocks with branching in continuum, involving a very large number of building blocks. As branching is a universal phenomenon, a comparison with other functional materials can be instructive. From a general point of view, branching occurs when a phase transformation results in the formation of different entities (like domains or variants). A microstructure consisting of these entities can undergo refinement when approaching a phase boundary. This saves volume energy at cost of interface energy.

Branching was first proposed by Landau for the refinement of intermediate states in type-I superconductors under applied field [19]. Here, the entities are represented by normal and superconducting volumes in the sample. Subsequently, the idea was developed for anisotropic ferromagnets by Lifshitz [20] and others [21]. The different entities are identified as different magnetic domains, e.g., up- and down-domains in a uniaxial easy-axis magnet. At a phase boundary to a non-magnetic region the (volume) stray-field energy is reduced on expense of increased (interface) domain wall energy. A detailed theory was worked out by Hubert [22], treating these various cases in a unified simple framework. Kohn and Müller [6] adopted these ideas to describe branching in martensite within a basic model that allows rigorous mathematical treatment, e.g. recent work [23].

Although in all these materials the entities consist of discrete building blocks (martensitic unit cells, spins, quanta), it remains open, if the present concept of coarsening of discrete building blocks is observable in other cases. In the martensite microstructure exmined here, only three orientations of a tetragonal martensitic unit cell are possible and the boundaries between these entities are atomically sharp. This suggests that, e.g. in magnets huge magnetocrystalline anisotropy may be required to observe a similar discrete coarsening process. It may be more feasible to examine ferroelectrics, which often exhibit a huge anisotropy, reducing the domain wall width to the atomic scale.

In all cases the compromise between volume and interface energies results in a universal scaling law describing the entity width $\Lambda$ in dependence of distance $x$ to the phase boundary: $\Lambda \sim x^{2/3}$. This suggests that the branched microstructure can be viewed as a fractal object, and for the Kohn-Müller model of martensites [6] it has been shown that energy minimizing solutions are asymptotically self-similar [23].

Along different lines, Hornbogen suggested to use fractal geometry to describe martensitic microstructures by applying the generating rules for a refinement [8,24]. Thus, in the present case, we may consider the similarity dimension $D_s$ (see e.g. [25,26]) of the branching twins, consisting of $N = 2$ segments (= two twin variants) and size scaling by a factor $n = 1/2$, which yields $D_s = \log N/\log (1/n) = 1$. Assuming that the twinned structure consists of homogeneous lamellae in the third dimension along the habit plane, we may conclude that the branched structure of the phase boundary has a fractal dimension of 2. This value indicates that branched martensites could be "borderline fractals" [26].

For a more detailed characterization the actual geometrical scales of the branching structure have to be considered. While the variant width Λ increases with distance to the habit plane by a factor of two with each generation, also their length increases with each coarsening step. This results in a different scaling along the habit plane compared to the perpendicular direction. Hence the observed branched martensite has to be described as self-affine fractal [27], which does not own a unique broken dimensionality according to standard definitions. In fact, the spontaneous strain $u \propto 1-c_{NM}/a_{NM}$ in the Kohn-Müller model [6], scales as function of the coordinates $(x,y)$ along and perpendicular to the habit plane as $u(x,y) \to \theta^{-2/3} u(\theta x, \theta^{2/3} y)$ with $\theta$ an arbitrary scale factor [23]. Therefore, we may consider the branching region at the habit plane as a fractal with an effective dimension between 2 and 3.

## 7. Symmetry breaking effects of a rigid substrate

When discussing the coarsening mechanism, we need to distinguish between issues relevant for the material itself (especially bulk Ni-Mn-Ga) and peculiarities of the thin film geometry. The present experiment crucially benefits from the thin film geometry since close to the substrate-film interface the martensitic transformation is suppressed. Coexistence of residual austenite and transformed martensite [17] requires the existence of a habit plane even well below the usual martensite finish temperature. As the elastic moduli of the MgO substrate strongly exceed those of austenite, the invariant length constraint is even stricter compared to bulk, in which some elastic deformation of austenite can occur. This can explain why for thin films the ratio of $d_1/d_2$ is fixed more accurately compared to bulk [12]. Additionally, for bulk material, the variant width is expected to increase uniformly with the distance from the habit plane [6]. In the present thin film experiment, however, various different generations are observed at a constant film thickness. This difference very likely originates from the different nucleation behaviour of martensite in the film. The relatively high energy of the habit plane may inhibit an easy nucleation of martensite. Thus, in bulk Ni-Mn-Ga, commonly only one (or a few) habit planes propagate through the entire single crystal. This is different for thin film. Due to substrate constraint and significantly lower film thickness compared to lateral extension, the different regions in the film are elastically decoupled resulting in large amount of nucleation sites. [14].

The two types of pattern observed on the sample surface indicate that from all these differently aligned habit planes only some allow a coarsening of the nanotwinned (14M) martensite up to macroscopic NM variants. In the present sample, we observed only two alignments of NM twin boundaries, which both evolve by coarsening of 14M variants with $c_{14M}$ (=$a_{NM}$) pointing out-of-plane. A microstructure of variants originating from coarsening of 14M variants with $b_{14M}$ and $a_{14M}$ pointing out-of-plane should exhibit twin boundaries with 45° rotated traces on the film surface, but no large scale features with this orientation have been observed. The absence of these types of macroscopic NM twin boundaries can be explained by substrate constraints as follows. The formation of a NM twin boundary results in a bending angle α = 90° - 2 arctan$(a_{NM}/c_{NM})$ = 11.6° between the crystal axes of both variants. If the twin boundary is inclined by 45° towards the substrate the $c_{NM}$-axis is alternating perpendicular and parallel to film plane. As sketched in Fig. 7, this formation of macroscopic NM twin boundary would require a significant bending of the rigid substrate or the formation of a gap. Since neither is possible, only the shortest twinning period of this type is formed ( = 14M) and no coarsening occurs. The observed alignment of $a_{NM}$-$c_{NM}$ twin boundaries perpendicular to the substrate surface does not require bending of the substrate. Formation of this type of twin boundary solely changes the orientation of the crystal axes within the film plane by 11.6° when passing the twin boundary. This process does not require surface buckling. This is consistent with surface topography where no pronounced height contrast is observed for these features (Fig. 4 (a) and red profile in (b)). Detailed look however, reveals that the NM areas also exhibit a small height contrast with long wavelengths up to the micrometre scale. This may be due to the fact that the 14M variants with $c_{14M}$ oriented out-of-plane, from which these macroscopic NM variants originate, are slightly tilted by ~2° towards the substrate normal [12]. Hence the twinning angle $\alpha$ is not perfectly in the film plane.

These symmetry breaking effects can explain why we observe 14M and NM coexisting at the film surface, a behaviour which would not be expected for bulk single crystals.

## 8. Transformation sequence and thermal hysteresis

The present results obtained from epitaxial films can provide a general explanation for peculiarities in the transformation behaviour and hysteresis of martensite bulk samples in general and magnetic shape memory alloys in particular. During cooling of Ni-Mn-Ga, often the sequence of austenite (A) - 14M martensite - NM martensite is observed [28]. This transformation sequence cannot be attributed to a simple (inter-) martensitic transition. During a transition from 14M to NM the crystal symmetry increases from (pseudo-) orthorhombic to tetragonal, which excludes a martensitic transition, in which the symmetry is reduced in the low temperature phase [11]. However, this apparent inconsistency can be explained considering that 14M-NM transition is not a phase transition but only a change of the martensitic microstructure as explained above. This argument is also supported by the small released heat during transition, which is about one tenth of latent heat of a common martensite transformation [29].

In the following we describe why the formation of an intermediate, adaptive structure like 14M is favourable for a transition to a tetragonal martensite. In order to obtain a reversible thermoelastic transition, the specific volume between austenite and martensite should be similar, which is satisfied for Ni-Mn-Ga. However, with constant volume, there is no possibility to form an exact habit plane between austenite and a single variant of the tetragonal martensite [11]. Consequently, the interface energy between austenite and martensite is relatively high, which commonly results in a huge thermal hysteresis. For an adaptive phase this is different since the exact habit plane is realized by the nanotwinning as described in Chapter 2. The existence of an exact invariant plane between austenite and martensite commonly results in a low thermal hysteresis [30, 31], as also observed for the A-14M transition in Ni-Mn-Ga [4].

Above we described how the tetragonal martensite forms an adaptive microstructure (= adaptive martensite) at the habit plane in the first stage of transition. In the next stage, the coarsening of twin variants occurs since this process reduces the overall twin boundary energy. In contrast to the thin film geometry, where a complete transformation to the tetragonal martensite is hindered by the constraint of the rigid substrate, in bulk the intermediate, adaptive martensite can disappear and the sample transforms to the ground state of tetragonal NM martensite [32, 33].

Commonly, the hysteresis during a structural transition is ascribed to a barrier imposed by the excess energy of defects that connect parent and product phase [34]. In the commonly found type of first-order phase transitions, these are interface energies caused by the appearance of phase boundaries. For the present case of a cubic to tetragonal martensitic transition, we can extend this idea. The hysteresis has its origin in two different types of defected microstructures. The hysteresis of the transition between austenite and adaptive phase is narrow, as expected from the low interface energy of the exact habit plane. It is not zero, as still small deformations at the habit plane may occur (gray plane marked in Fig. 1 b). This first, small contribution to hysteresis is induced by the "classical" interface energy derived from the habit plane. The major part of the whole hysteresis for an A-NM transition originates from the fractal interface forming during the coarsening process, which is analysed in the following.

At atomic scale the coarsening requires an unfavourable bending of twin boundaries, which can be realized by glide and climb of disconnections as described in [35], and the annihilation of two twin boundaries of opposite sign. The rationale for these processes is evident from the continuum picture (Fig. 2). On the atomic scale both processes are associated with a significant disturbance of the lattice (as an illustration, one may try to connect the $0^{th}$ and $1^{st}$ generation in Fig. 1). The coarsening process starts at the habit plane and proceeds into the martensite. Hence the associated defect energies are spatially distributed over a broad region. Following the continuum model, long range elastic coupling tends to keep the ratio between both variants constant at a certain distance to the habit plane (Fig. 2). This makes the whole branched arrangement of twins stiff. Thus all annihilation processes should occur simultaneously at well defined distances from the habit plane. Indeed the observed large regions with a regular twin pattern (chapter 5) suggest that coarsening is a collective process involving many twin boundaries. Though the energy barrier for local motion of an individual twin boundary (under the driving energy of a coarsening process) might be overcome by thermal activation, the coarsening process requires a collective rearrangement of the branched microstructure. As a non-local process the energy barrier for coarsening is a multiple of the individual processes. The energy barrier for coarsening may increase up to a range which cannot be overcome anymore by thermal activation. During the first stage of coarsening (from $n = 0$ to 1) the highest absolute number of twin boundaries is annihilated. Hence this energy barrier is expected to be largest compared to the following coarsening stages. This barrier allows retaining modulated martensite as a metastable phase in a broad temperature range. This implies that modulated phases should not be observable at higher temperatures. Indeed for the Ni-Mn-Ga system only tetragonal structures are reported for compositions exhibiting a martensitic transformation temperature well above room temperature [36].

This scenario can explain the relatively large hysteresis of a complete cubic to tetragonal (NM) martensitic transition in Ni-Mn-Ga alloys compared to the A-14M transformation [29]. In contrast to equilibrium phases, for a metastable phase one has to distinguish between driving forces for a phase transition and the energy barrier hindering the transformation. In case of a transformation from 14M to NM the twin boundary energy represents the driving energy for coarsening while the microstructural defects associated with coarsening are the energy barrier stabilizing the metastable adaptive phase. This may justify describing modulated martensite as adaptive "phase" and not microstructure. Metastability allows utilizing single crystals of the adaptive modulated phase at room temperature for several $10^8$ actuation cycles [37].

While commonly martensitic transformations are considered as athermal [38,39], we predict that thermal activation should be important for the transition between the metastable, adaptive phase and the tetragonal ground state. We expect that detailed time and cooling rate dependent measurement will confirm our concept.

## 9. A hierarchically twinned microstructure

In the following we analyse the microstructure that forms when a high energy barrier efficiently hinders coarsening. In this case one may consider a 14M unit cell as a mesoscopic building block for the martensitic microstructure. Due to its adaptive origin, it compensates the linear elastic strain at the habit plane. An equivalent formulation is that it can form an exact habit plane (Chapter 2). From this point of view, no further twinning is required to fully transform the austenite through an exact habit-plane into one 14M variant. However in the bulk of the material, the transformation also causes strains in directions perpendicular to the habit plane, which causes stresses in the surrounding material. This stress increases with the size of the 14M variant. At a certain extension it becomes more favourable to form twin boundaries with a differently aligned 14M variant. In order to distinguish this type of twin boundary from the primary (common) twin boundary connecting tetragonal variants, we will call them mesoscopic twin boundaries.

In our experiments on the film these mesoscopic twin boundaries have a spacing of about 80 nm, which is about 50 times larger than $\Lambda_0$, representing the primary twinning period of the adaptive martensite. This indicates that a large amount of elastic energy due to shear deformation has accumulated. As with primary twinning, the formation of energetically unfavourable mesoscopic twin boundaries can reduce volume elastic energy. The larger spacings observed for mesoscopic twin boundaries suggest that they presumably have a higher energy compared to the primary twin boundaries of the tetragonal martensite. This is revealed by the significant curvature observed for mesoscopic twin boundaries (Fig. 4 and 6) and the higher mechanical stress required to nucleate mesoscopic twin boundaries in a single crystal [40, 41].

Although mesoscopic twin boundaries release most of the deformation a small deformation remains even for a compound of mesoscopic variants. This can result in a further hierarchy of twins with much larger spacing. The role of these "macrotwins" is not understood to date. In Fig. 6 a sketch of Roitburd [7] is shown, which illustrates this process.

Due to the rotation at a nanotwin boundary some shear with the principal axes along the diagonals of the austenite remains. Part of this elastic energy can be diminished if the a next generation of twins is introduced in a hierarchy with boundaries that are rotated by about 45° with respect to the previous generation of boundaries. For comparison experimentally observed twin patterns on an epitaxial Ni-Mn-Ga film for the 3 different relevant length scales are shown. As the coarsening process of the present film disturbs the regular pattern, the micrographs of a film consisting only of 14M are taken from [14]. From the present experiments the first level of twinning can only be deduced from XRD [12], but recently scanning tunnel microscopy also revealed a direct image of a 14M modulation at atomic scale on epitaxial Ni-Mn-Ga films [42]. While the second level of mesoscopic twinning has a periodicity of about 80 nm, the third, self-similar hierarchy ranges up to 0.1 mm.

Though there are some reports on hierarchical twinning in bulk Ni-Mn-Ga [16,43] these patterns are often disturbed by the third variant orientation possible. For a thin film we benefit from the variant selection described in chapter 7. Indeed the sketch used for comparison (Fig. 6) is for a tetragonal – orthorhombic transition, allowing only two variant orientations [7].

## 10. Mobility of mesoscopic of twin boundaries

A key question for the existence of magnetically induced reorientation (MIR) remains unanswered; which type of twin boundary is highly mobile and why? To elucidate this we analyse the differences between 1) primary (common) twin boundaries connecting tetragonal martensite and 2) mesoscopic twin boundaries connecting variants of the adaptive 14M phase.

The formation of a nanotwinned adaptive martensite requires a high density of primary twin boundaries connecting differently oriented NM martensite variants. These twin boundaries are highly symmetric and atomically sharp as indicated by high resolution microscopy [43]. They exhibit very low twin boundary energy [12]. The crystallography of these nanotwin boundaries is identical to twin boundaries in macroscopic NM crystals. In some regions disconnections can be found and their movement is believed to be a microscopic mechanism for twin boundary motion [44]. Stress-strain measurements of NM single crystals, however, reveal that this kind of twin boundary is not mobile enough to be moved by an external magnetic field [45,46], which is in agreement with ab-inito calculations [47]. Though recently a small strain induced by a magnetic field had been reported for an NM single crystal [48], the twinning stress of NM martensite is far above the 0.05 MPa recently obtained in modulated martensite [49]. Crucially Soolshenko at al.[50] reported a large difference in twin boundary mobility in NM and 14M martensites measured in an identical sample and the same temperature range. These observations seem to exclude a different defect density as a possible origin of different twin boundary mobility. Instead these experiments suggest that a fundamentally different type of twin boundary with high mobility exists in the modulated martensite.

High mobility is observed e.g. for $a_{14M}$-$c_{14M}$ twin boundaries [4, 46]. In order to obtain an illustrative understanding of the geometry of such a twin boundary, we use the common approach to construct a twin boundary by taking a unit cell and its mirrored counterpart and connect them. In Fig. 8, a photograph is shown which illustrates the incompatibility of this mesoscopic twin boundary (a foldable 3D-model is added in the supplementary material). This is in contrast to the sharp twin boundaries between NM variants, having a very low energy [12]. The large monoclinic 14M crystal unit cell does not allow the formation of a simple $a_{14M} - c_{14M}$ twin boundary with most of the atoms joining both unit cells. A rigid crystal would leave a gap between both variants. Müllner and Kostorz [44] realized that a large amount of dislocations is required to form an interface between 14M variants. They suggested that some rearrangements can reduce the misfit, but still the incompatible twin boundary is expected to have an unfavourable high energy – in agreement with the consideration in Chapter 8. Han et al. [51, 52] examined mesoscopic twin boundaries by TEM. Though they describe these boundaries as rather flat on the nanometer scale, the micrographs reveal fringes with a thickness of up to several nanometres. They suggest that these fringes are caused by shuffling and deshuffling of atoms in the vicinity of twin boundary in order to reduce elastic energy. This feature supports our assumption of a diffuse boundary on the atomic scale. Indeed when shifting the 3D models slightly towards each other, the incompatibility can be reduced. As a consequence, however, this boundary maintains neither mirror nor rotational symmetry. High resolution micrographs reveal that this type of boundary exhibits no inversion symmetry [51]. However, from a macroscopic point-of-view, they have all properties of a twin boundary. As a consequence of the symmetry operation described by continuum theory, a mesoscopic twin boundary does not necessarily exhibit a simple structure at the atomic scale, (see [11], page 68 for a discussion on the different definitions of twin boundaries used in different communities). This justifies calling them mesoscopic twin boundaries.

Though recently Müllner and King [53] made an attempt to expand the dislocation/disclination approach for the movement of mesoscopic twin boundaries, we suggest that the origin of the extraordinary mobility of these mesoscopic twin boundaries arises from their diffuse nature compared to the atomically sharp interface of a common twin boundary.

The mobility of twin boundaries can be hindered by various types of defects e.g. point defects, dislocations, chemical disorder, antiphase boundary, precipitates, which even exist in samples close to perfect single crystals. For an atomically sharp twin boundary very small point defects can act as efficient pinning centres. When defect extension is comparable to the width of a twin boundary, a high force may be required to detach them. Since all Ni-Mn-Ga alloys exhibiting MIR at room temperature are non-stoichiometric, the high number of site disorder within the Heusler lattice may be sufficient for an efficient pinning of atomically sharp twin boundaries.

We expect that the situation is different for a mesoscopic, diffuse twin boundary in which the boundary structure is not sharply localized within a few atomic planes but it is distributed over several nanometres. Therefore, the defect energy densities of a mesoscopic twin boundary are smoothed on the scale of many tens of lattice spacings. Pinning forces depend on the gradient of the defect energy density hence a broad twin boundary is expected to be only weakly pinned by small defects. Only larger defects, as precipitates or voids, can pin the twin boundary efficiently [54]. Additionally, the diffuse nature of the mesoscopic twin boundaries might also permit certain deviations from a fixed (101) orientation which means that these twin boundaries could bend and adjust to defects more easily.

At the first sight it appears unlikely that the movement of a diffuse mesoscopic twin boundary allows restoring a complex, metastable structure as 14M. However, during the movement the macroscopic extension of the diffuse twin boundary does not change. Thus it represents an invariant plane – in analogy to a habit plane. As a consequence of this constraint, the average $d_1/d_2$ ratio cannot change. Moreover, mechanically or magnetic field induced, repetitive movement of a mesoscopic twin boundary may smooth the local fluctuations of $d_1$ and $d_2$, which can reduce the number of stacking faults in the nanotwinned lattice. We speculate that this is the microscopic mechanism for training, which is known to increase the twin boundary mobility during several mechanical or magnetic cycles [55,56].

More detailed microstructural investigations are required to confirm these concepts. We suggest using isostructural, but non-magnetic martensitic materials for these experiments. The magnetic field created by the field lens in a TEM is in the Tesla range and therefore sufficient to move twin boundaries in Ni-Mn-Ga. Hence if one observes a mesoscopic twin boundary it is clear that this boundary must be efficiently pinned. A mobile twin boundary would move and disappear in the lens field and it is therefore not observable in a usual set-up.

# 11. Modulated martensites beyond Ni-Mn-Ga

In the present experiments for Ni-Mn-Ga we show how modulated martensite can form due to the geometrical constraints at the habit plane. As this is a typical feature of diffusionless transformations, modulated structures can be expected for a broad range of materials. First of all, for the Ni-Mn-Ga system they are not limited to the exemplary case 14M examined here but our previous analysis shows that the adaptive concept can also explain 10M and premartensite (6M) modulations [12]. $Fe_{70}Pd_{30}$ as the second magnetic shape memory alloy discovered [57] also exhibits modulated structures [58], identified as adaptive phases [5].

In addition to these metallic alloys, adaptive phases are often found in ferroelectrics like PMN-PT [59]. These ferroelectrics can reach strains well above one percent [60]. While the microscopic actuation mechanism is similar to MIR, electric fields are used instead of magnetic fields to move twin boundaries [61]. Modulated structures form in a transitional region in vicinity of a morphotropic phase boundary [59,62,63]. Since there are difficulties to describe them as equilibrium phases, it has been suggested to explain the anomalous phenomena at the morphotropic phase boundary by bridging structures [64]. Furthermore, recently a hierarchical [65] and fractal [66] twin microstructure had been observed.

These similarities with the metastable modulated phases in Ni-Mn-Ga establish magnetic shape-memory alloys as an important metallic counterpart to ferroelectrics near the morphotropic phase boundary. The formation of an adaptive phase seems to be crucial for field-induced giant strains in martensitic functional materials. Modulated phases facilitate adaptation to external forces and fields by a redistribution of mesoscopic twin boundaries, in contrast to a thermodynamically stable, stiff martensite.

It is worth to add that already in 1974 Anderson and Hyde [67] identified twinning at the unit cell level as a concept for structure building. Materials include currently intensively examined materials like $Fe_3C$, $BaTi_4O_9$, $Fe_2TiO_5$. To date such crystallographic structures are known in a broad range of materials [68]. Hence, modulated, seemingly thermodynamically stable phases in various other materials may find a consistent explanation as metastable adaptive microstructures on closer scrutiny. Moreover, the formation of such modulated structures may not necessarily be restricted to diffusionless transformations, but it could arise as a general feature of phase transformations under the transient constraint of a phase boundary in elastic solids.

# 12. Conclusions

The presented experiments allow sketching relations between the martensitic microstructure and the extraordinary functional properties of modulated martensite. We illustrate how the macroscopic geometrical constraints of a habit plane can act down to the atomic scale, resulting in the formation of a nanotwinned, adaptive phase. Within this concept the modulated structure is only determined by the lattice constants of its tetragonal building blocks and the lattice constant of the austenite. The concept explains the particular relation between the lattice constant of the modulated and non-modulated martensite and the austenite. Combined with a coarsening mechanism, the observation of the self-similar microstructure of the non-modulated martensite can be understood. Here, the finite size of the nanotwinned variants is the starting point for the coarsening process of martensitic variants, which tends to produce macroscopic variants of the tetragonal martensite as ground state. Coarsening reduces the amount of twin boundary energy by doubling variant widths, while keeping their width ratio fixed (Fig. 5). This constitutes a fractal process which could be followed for 10 generations from the atomic up to the micrometer scale. We suggest that the coarsening mechanism can close the gap between atomic description and continuum theory.

If one is to accept the adaptive concept and the picture of a branched microstructure that coarsens the martensite, one has to explain the energy barrier that stabilizes the modulated structure as a thermodynamically metastable state. We have described how this barrier can be associated with a collective coarsening process as the elastic constraint hinders the selective growth of single nanotwins. This energy barrier contributes to thermal hysteresis and selects between two different microstructures:

1) If the energy barrier can be overcome, a macroscopically twinned, tetragonal martensite forms. We attribute the relatively low mobility of twin boundaries in tetragonal martensites to their high symmetry, which results in an atomically sharp boundary that is easily pinned by point-like lattice defects.

2) If the energy barrier cannot be overcome, a self similar hierarchical microstructure exhibiting "twins within twins" forms. This idea explains why in the experiments reported here three different levels of twins from the nanometre to the millimetre scale are observed. The first level of twins can be described as a metastable adaptive "phase". At the second level the mesoscopic twin boundaries between variants of the adaptive phase may form. As these mesoscopic twin boundaries are to be seen as interfaces between structures that are not thermodynamically stable, these interfaces are expected to differ from common atomically sharp twin boundaries between elastically stiff phases. Thus, we suggest that these mesoscopic twin boundaries have a diffuse nature and a high boundary energy. Then, we can attribute the macroscopically observable high mobility of these mesoscopic twin boundaries to a broad pinning potential, which allows an easy glide over small lattice defects.

In the present epitaxial film both types of microstructures were observed at the same time. This behaviour can be attributed to the constraints of the rigid substrate, which inhibits bending of the film and a macroscopic length change. The existence of these features permits a detailed analysis of the martensitic microstructure.

**Acknowledgements**

The authors acknowledge the fruitful discussions with Manfred Wuttig, the help of Svea Fleischer for preparation of the manuscript and the instructive textbook of Kausik Bhattacharya, which inspired the present article's title. This work receives financial support from the Priority program SPP 1239 (fund number FA 453/7 and RO 2238/8) (www.MagneticShape.de), the project FA 453/8 from the German Research Society and from the Academy of Sciences of Czech Republic.**References**

**Figure captions**

Fig. 1: *(color online) (a) The unit cell of cubic austenite can transform into three equivalent orientations of a tetragonal martensite unit cell. (b) Sketch of the orientation relationship between parent austenite and nanotwinned (adaptive) martensite phase. The different blue background colours mark differently oriented tetragonal martensitic variants, which are connected by twin boundaries. The gray plane marks the habit plane. It is drawn with finite thickness in order to illustrate that it is accompanied by a distortion of the lattice. (c) Since the high density of twin boundaries is ultimately energetically unfavourable, coarsening of tetragonal twin variants may occur by annihilation of twin boundaries. Exemplarily the first generation of coarsening is shown, where the period doubled compared to the nanotwinned martensite (d) shows a macrotwinned martensite where the twin boundaries have a macroscopic distance compared to the atomic distances shown in (a) to (c).*
*The generation n of coarsening reefer to the analysis of the coarsening process described in chapter 5.*

Fig. 2: *(color online) Sketch of the continuum model describing branching of twin boundaries connecting two different variant orientations (blue and green) when approaching the habit plane towards austenite (red) (adapted from [6]). The elastic energy originating from the different lattice constants between austenite and martensite is reduced on cost of increased twin boundary energy. During this process the fraction of both variants is expected to remain constant.*

Fig. 3: *(color online) Micrographs of an epitaxial Ni-Mn-Ga film used to analyse the microstructure and twin boundary orientations of 14M (green) and NM (red) martensite. The picture edges are parallel to MgO <100> directions.*
*(a) Large scale SEM micrograph of the film surface revealing areas with a stripe-like contrast parallel to the substrate edges and areas with a fine morphology that seems irregular at this magnification. When zooming in the picture (b) again the regular patterns parallel to the substrate edges appear (red line). In addition a finely twinned microstructure in some parts of the film, rotated 45° with respect to the substrate edges (green line) becomes visible. (c) Top part of this figure shows the film surface and the bottom part a cross section through the film as prepared by FIB. The substrate underneath is not shown. (d) Orientation of the Ni-Mn-Ga unit cells epitaxially grown on the MgO substrate. The different orientations of twin boundaries are coloured corresponding to the features marked in the micrographs.*

Fig. 4*: (color online) (a) AFM micrograph of the surface topography revealing the same characteristic two types of features observed in SEM. Finely twinned regions (green line) exhibit higher height contrast while the large features (red) are almost flat. In (b) the height profiles along both lines marked in (a) are plotted. Additionally, the characteristic twinning angle $\alpha$ is sketched in the height profile of 14M. Picture edges are parallel to MgO <100> directions.*

Fig. 5: *(color online) Analysis of variant widths $d_1$ and $d_2$ obtained in various regions on the sample surface. The x-axis corresponds to the generation of coarsening due to doubling of variant size. The widths obtained by XRD are assigned to the $0^{th}$ generation (open symbols). The variant ratio $d_1/d_2$ (red dots) remains constant while the width $\Lambda = d_1 + d_2$ (black squares) extends up to the micrometer range. All observed different variant widths can be assigned to an integer generation n of coarsening ($\Lambda_n = 2^n * \Lambda_0$, black line).*

Fig. 6: *(color online) Formation of a hierarchically twinned microstructure from the atomic (left) to macroscale (right). The top row sketches the twin boundary orientation expected from continuum theory (adapted from [7]). The bottom row starts with a sketch of the 14M modulated martensite as derived from XRD and then SEM micrographs of an epitaxial 14M Ni-Mn-Ga film are shown (taken from [14]). The overall microstructure reflects the expected 4-folded symmetry of*

*the substrate. For clarity, only regions with a 2-folded symmetry are depicted. The edges of these graphs are parallel to the austenitic unit cell (and hence rotated 45° compared to Fig. 3 and 4).*

Fig. 7: *(color online) Incompatibility of a twin boundary angle $\alpha$ with a rigid substrate. The formation of a twin boundary in a film requires either bending of the substrate or the formation of a gap. Since neither is possible for a film attached to a rigid substrate without severe deformation, this orientation of macroscopic twin boundaries is disfavoured. The elastic energy associated with the deformation of the unit cells is minimal at the shortest variant length (= modulated structure of 14M).*

Fig. 8: *(color online) Illustration of a mesoscopic $a_{14M}$-$c_{14M}$ twin boundary using the foldable 3D paper model available as supplementary material. The gap between both parts illustrates the incompatibility, which originates from the complex 14M unit cells. Shifting both sides along $b_{14M}$ can reduce the gap partly and, together with lattice reconstruction, may be a mechanism to reduce the high energy of a mesoscopic twin boundary (not shown).*

Fig. 1

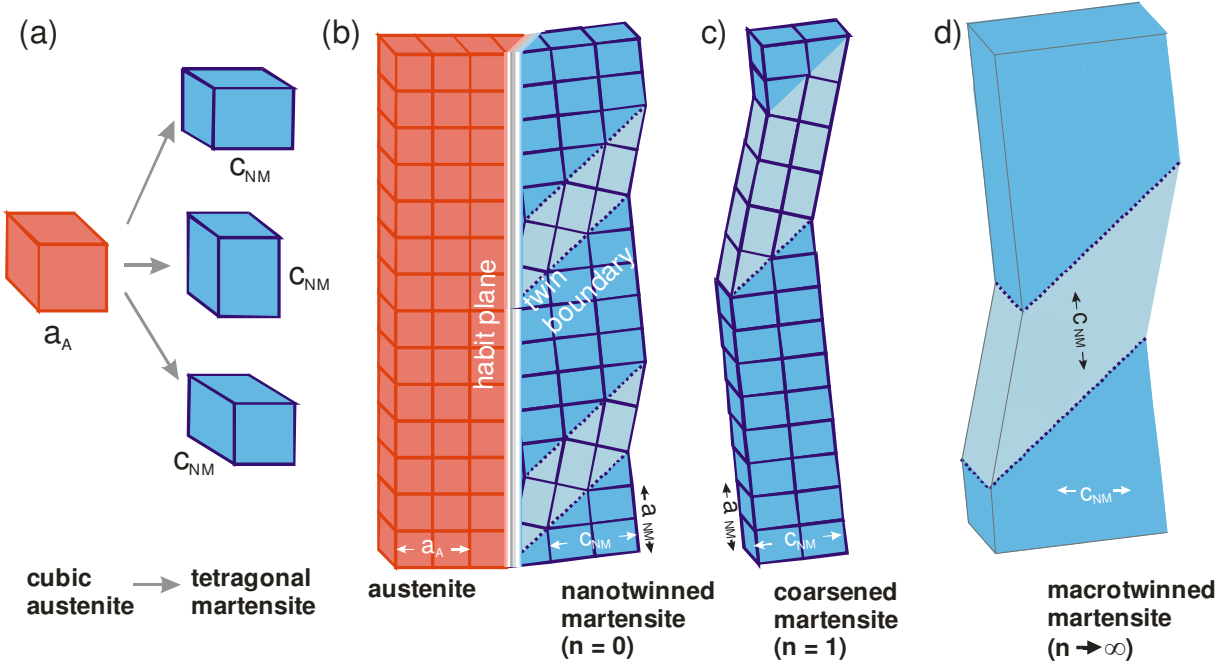

Fig. 2:

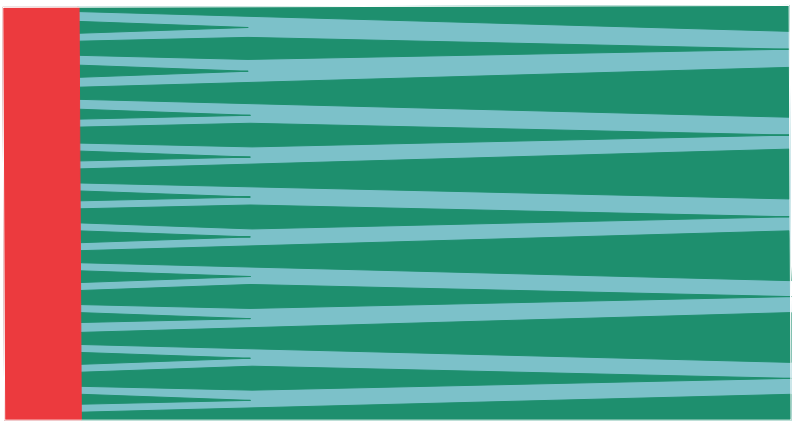

Fig. 3:

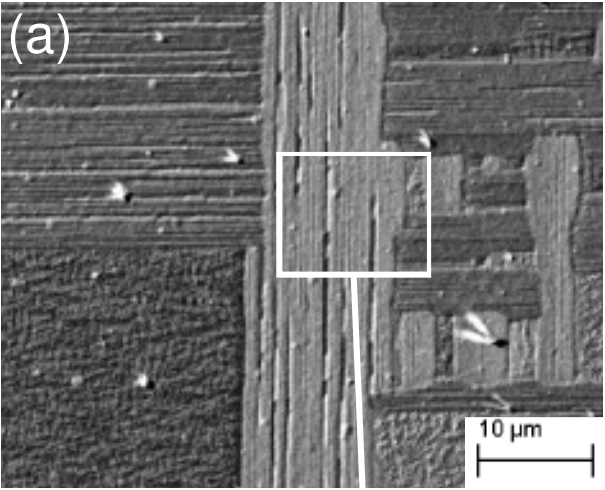

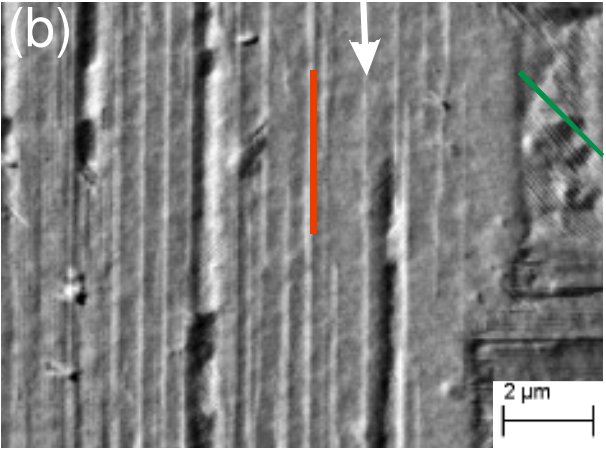

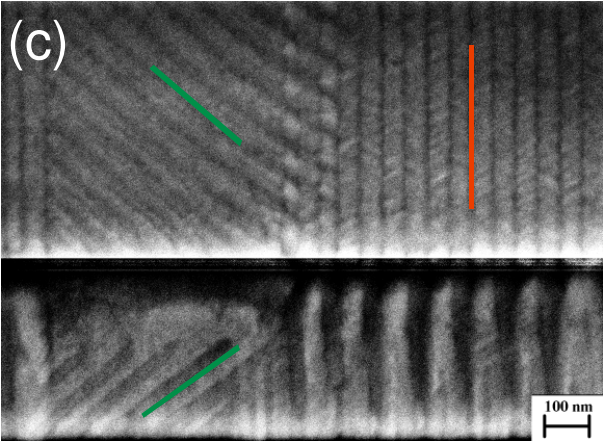

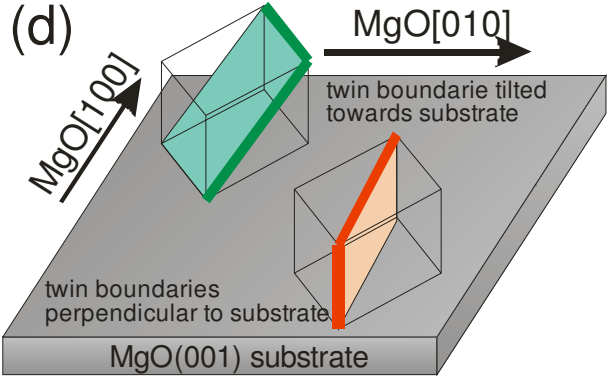

Fig. 4:

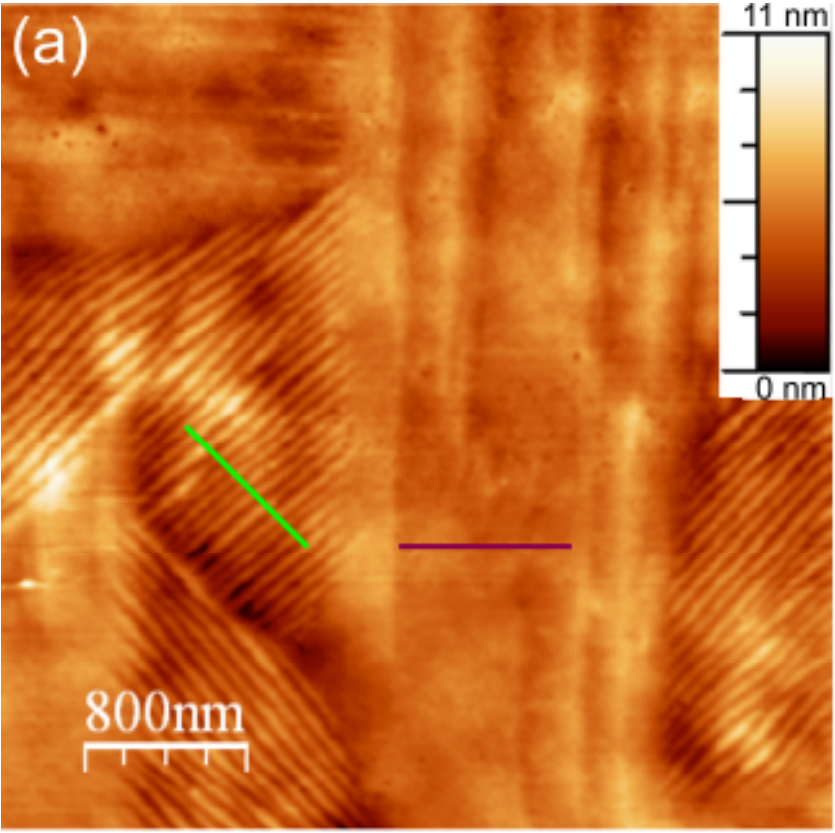

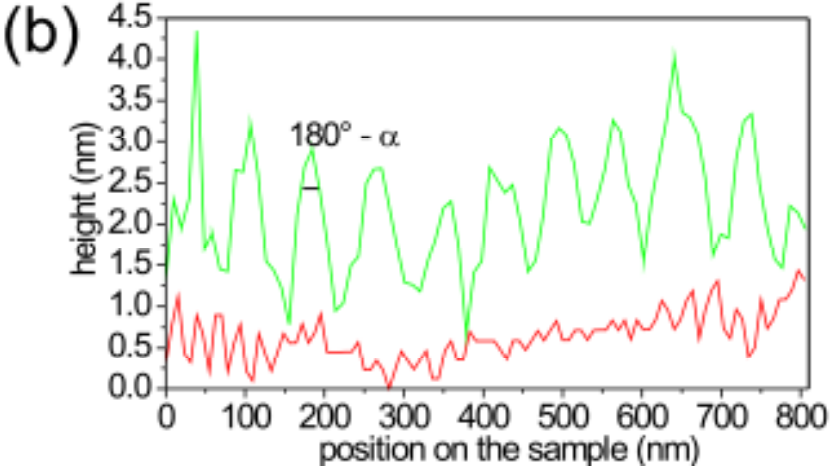

Fig. 5:

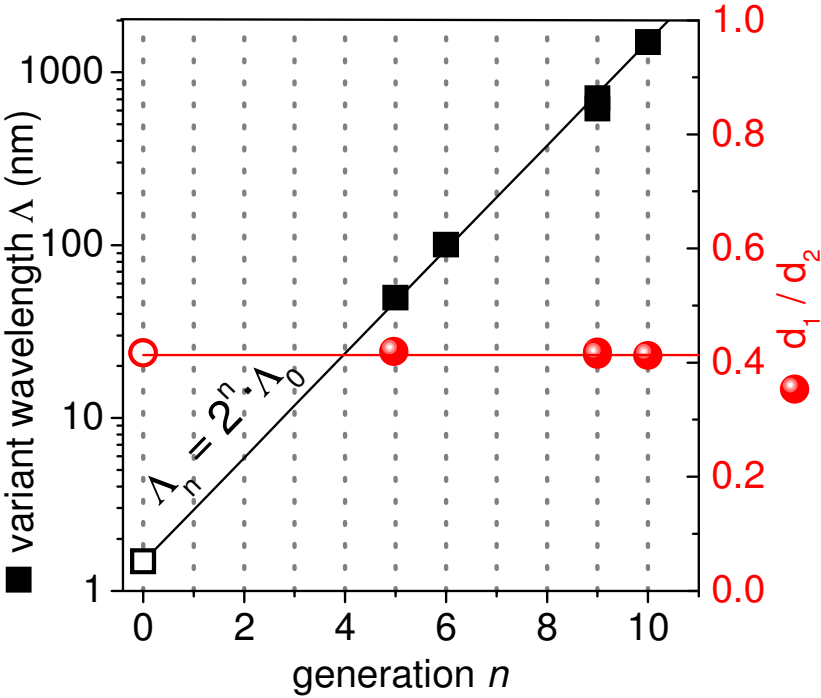

Fig. 6:

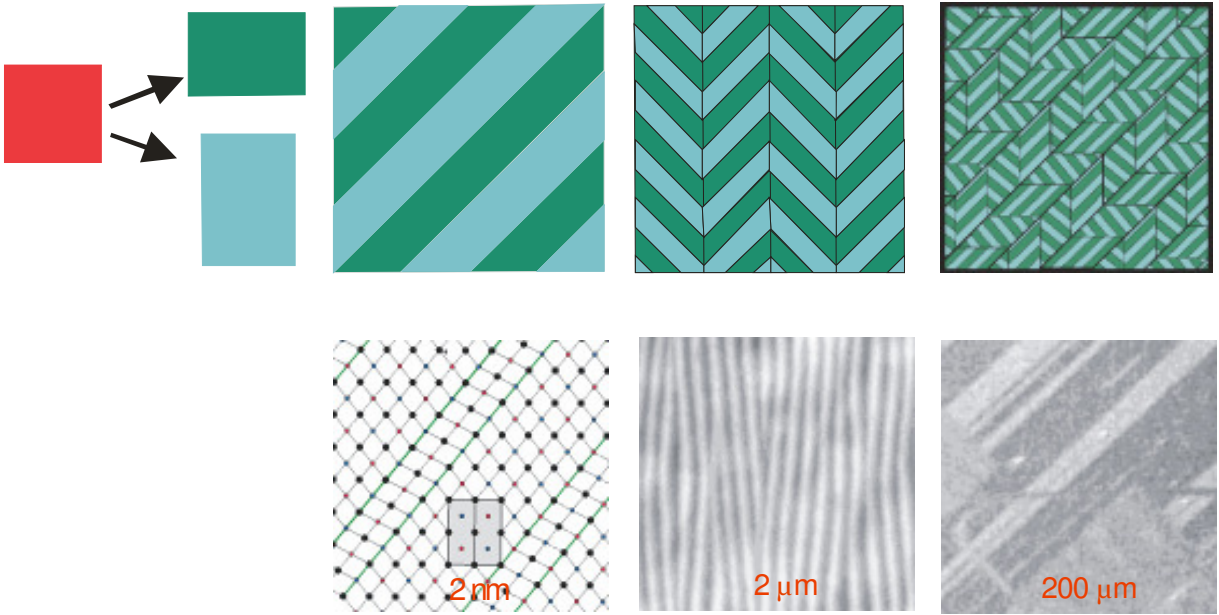

Fig. 7:

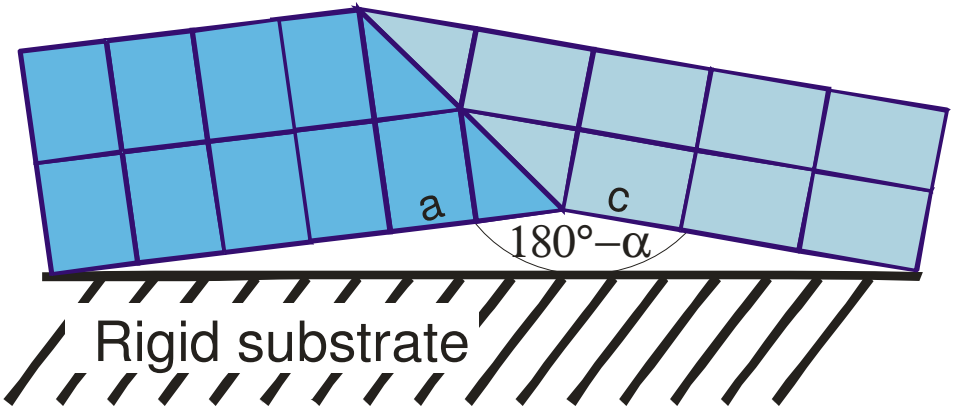

Fig. 8:

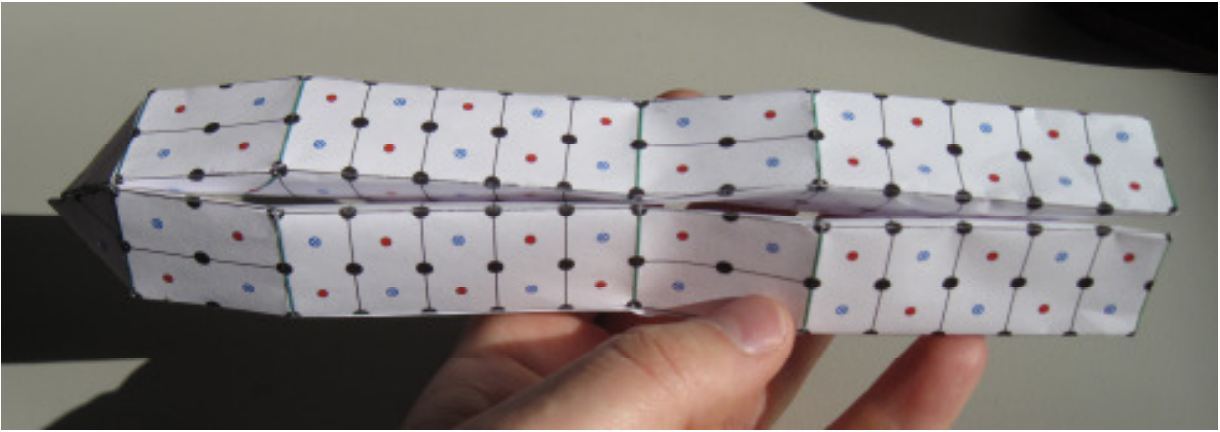